\documentclass{article}
\usepackage{multicol}
\usepackage{amsmath}
\usepackage{latexsym}

\newtheorem{lemma}{Lemma}[section]

\begin{document}

\title{Pattern Matching and Local Alignment 
for RNA Structures\footnote{Appeared in \emph{Proceedings of The 2002 International Conference on Mathematics
and Engineering Techniques in Medicine and Biological Sciences (METMBS), 55-61, 2002.}}}
\author
{
Shihyen Chen\footnote{Current Email: shihyen\_c@yahoo.ca},
Zhuozhi Wang, and
Kaizhong Zhang\\
Department of Computer Science\\
University of Western Ontario\\
London, Ontario N6A 5B7, Canada \\
\{schen, zzwang, kzhang\}@csd.uwo.ca
}
\date{}
\maketitle

\thispagestyle{empty}

\begin{multicols}{2}

\section*{Abstract}
\label{sec:abstract}
\emph{\noindent
The primary structure of a ribonucleic acid (RNA) 
molecule can be represented as a sequence of 
nucleotides (bases) over the alphabet 
$\left\{A, C, G, U\right\}$. 
The secondary or 
tertiary structure of an RNA is a set of base pairs
which form bonds between $A-U$ and $G-C$. 
For secondary structures, these bonds have been 
traditionally assumed to be one-to-one and 
non-crossing. 
\\
This paper considers pattern matching as well as
local alignment between two
RNA structures. 
For pattern matching, we present two algorithms, 
one for
obtaining an exact match, the other for approximate
match. 
We then present an algorithm for RNA local
structural alignment.
}

\section{Introduction} 
\label{sec:intro}
Ribonucleic Acid (RNA) is an important molecule which
performs a wide range of functions in biological systems.
In particular, it is RNA (not DNA) that contains genetic
information such as HIV and regulates the functions of
such virus. 
RNA has recently become the center of much
attention due to its catalytic properties, leading to
an increasing interest in obtaining its structural 
information.
\par
It is well known that secondary and tertiary structural 
features of RNAs are important in the molecular mechanisms
involving their functions. 
The presumption is that, to a 
preserved function there corresponds a preserved molecular
conformation and, therefore, a preserved secondary and 
tertiary structure. 
Hence the ability to compare RNA
structures is 
useful \cite{BMR95,CM95,E99,JLMZ00,LRV98,ZhangWangMa}.
In many problems involving RNAs \cite{CLM00,SBHMSUH94},
it is required to have an alignment between RNA structures
in addition to a similarity measure \cite{S85}.
In RNA secondary or tertiary
structure, a bonded pair of bases (base pair) is usually
represented as an edge between the two complementary bases
involved in the bond. 
It is assumed that any base 
participate in at most one such pair. 
For secondary 
structures, the edges of the bonded pairs are non-crossing.
\par
In \cite{ZhangWangMa}, a distance measure
between two RNA structures is proposed. 
This measure takes
into account the primary, the secondary and the tertiary
information of RNA structures. 
This measure treats a base
pair as a unit and does not allow it to match to two
unpaired bases. 
In general this is a reasonable model since,
in RNA structures, when one base of a pair changes, we
usually find that its partner also changes so as to conserve 
that base pair. 
Computing edit distance for tertiary 
structures is proved to be NP-hard \cite{ZhangWangMa}.
\par
In \cite{WangZhang}, a model based on the one defined 
in \cite{ZhangWangMa} is proposed to take affine gap 
penalty into consideraction. 
Experimental results show that the 
model in \cite{WangZhang} performs better than that 
in \cite{ZhangWangMa}. 
\par
In this paper, we consider a structural pattern matching
problem and a local structural alignment problem. 
\par
In the first problem, specifically, given two RNA 
structures $R_1$ and $R_2$, we want to find in $R_2$ a pattern
matched by $R_1$, with both exact match and approximate
match. 
For exact match, we present a labelling device
so that an RNA structural matching problem can be treated
by existing fast string-matching algorithms. 
For approximate match, we present an algorithm based on 
the model defined in \cite{WangZhang} with modifications
to adapt to the RNA structural pattern matching problem.
\par
In the second problem, specifically, given two RNA 
structures $R_1$ and $R_2$, we want to find the most
similar regions, one in each structure.
We present an algorithm also based on 
the model defined in \cite{WangZhang} with modifications
to adapt to the RNA local structural alignment problem.

\section{RNA Structural Representation} 
\label{sec:rep}
Given an RNA structure $R$, we use $R[i]$ to denote the 
$i$th base of an RNA structure $R$, 
and $R[i \cdots j]$ the sequence of bases from $R[i]$ to $R[j]$.
We use $|R|$ to denote the primary-sequence length of $R$.
A set of structural elements for $R$ may be represented by
$S(R)$, where $S(R) = U(R) \cup P(R)$, 
$
U(R) = \left\{i\ |\ R[i]\text{ is an unpaired base in }R\right\}
$ and
$P(R) = \left\{\left(i, j\right)\ |
\ i<j\text{ and }\left(R[i], R[j]\right)
\text{ is a base pair in }R\right\}$.
For $(i,j) \in P(R)$, $i$ is called the 5' end, 
and $j$ the 3' end.
We assume that base pairs in $R$ do not share 
participating bases.
Formally, for any $(i_1, j_1)$ and $(i_2, j_2)$ $\in$ $P(R)$, 
$j_1 \ne i_2$, $i_1 \ne j_2$,
and $i_1 = i_2$ if and only if $j_1 = j_2$.
We define $p(R[i])$ as \\
$
p(R[i]) =
\left\{
\begin{array}{l@{\quad}l}
i & \text{if } R[i] \in U(R), \\
j & \text{if } (R[i], R[j])\text{ or }(R[j], R[i]) \in P(R). 
\end{array}
\right.
$
\\
By this definition, $p(R[i])$ indicates if $R[i]$ is an
unpaired base (case 1) or a paired base (case 2) and
with which base it pairs.

\section{RNA Structural Alignment} 
\label{sec:align}
Following the tradition in sequence comparison
\cite{NeedlemanWunsch70, SmithWaterman81}, 
we define three edit operations on RNA structures: 
substitute, delete and insert.
For a given RNA structure $R$, each operation can
be applied to either an unpaired base or a base
pair. 
To substitute a base pair is to replace 
one base pair with another. 
This means that at the
sequence level, two bases may be changed at the same 
time. 
To delete a base pair is to remove the base pair.
At the sequence level, this means to delete two bases
at the same time. 
To insert a base pair is to insert a
new base pair. 
At the sequence level, this means to
insert two bases at the same time. There is no relabel
operation to change a base pair to an unpaired base or
vice versa. 
\par
We denote an edit operation by $a \rightarrow b$, where
$a$ and $b$ are either $\lambda$, the null label, or
labels of base pairs from $\{A, C, G, U\} \times 
\{A, C, G, U\}$,
or unpaired bases from $\{A, C, G, U\}$.
\par
We call $a \rightarrow b$ a substitute operation if 
$a \ne \lambda$ and $b \ne \lambda$;
a delete operation if $a \ne \lambda$ and $b = \lambda$; 
an insert operation if $a = \lambda$ and $b \ne \lambda$.
Let $\Gamma$ be a similarity-scoring function that
assigns to each edit operation $a \rightarrow b$ a
real number $\Gamma(a \rightarrow b)$. 
We constrain $\Gamma$ to be a similarity metric.
\par
Given two RNA structures $R_1$ and $R_2$,  
we adopt the following sign convention.
\\
\\
$
\begin{array}{l@{\,}l}
{} & \Gamma(R_1[i] \rightarrow R_2[j]) \ge 0
\text{ if } R_1[i] = R_2[j],\\
{} & \Gamma(R_1[i] \rightarrow \lambda) \le 0,
\Gamma(\lambda \rightarrow R_2[j]) \le 0,\\
{} & \Gamma((R_1[i_1], R_1[i_2]) 
\rightarrow (R_2[j_1], R_2[j_2])) \ge 0\\
{} & \qquad \text{if } R_1[i_1] = R_2[j_1] 
\text{ and } R_1[i_2] = R_2[j_2],\\
{} & \Gamma((R_1[i_1], R_1[i_2]) \rightarrow \lambda) \le 0,\\ 
{} & \Gamma(\lambda \rightarrow (R_2[j_1], R_2[j_2])) \le 0.
\end{array}
$
\\
\par
Given two RNA structures $R_1$ and $R_2$, a structural
alignment of $R_1$ and $R_2$ is represented by
$(R'_1,R'_2)$ satisfying the following conditions.

\begin{enumerate}
\item 
\label{cond:1}
$R'_1$ is $R_1$ with some new symbols $'-'$,
denoting a space, 
inserted and $R'_2$ is $R_2$ with some new symbols
$'-'$ inserted such that $|R'_1| = |R'_2|$.
\item 
\label{cond:2}
If $R'_1[i]$ is an unpaired base in $R'_1$,
then either $R'_2[i]$ is an unpaired base in $R'_2$
or $R'_2[i] = \ '-'$. 
If $R'_2[i]$ is an unpaired base in 
$R'_2$, then either $R'_1[i]$ is an unpaired base in $R'_1$
or $R'_1[i] = \ '-'$. 
\item 
\label{cond:3}
If $(R'_1[i], R'_1[j])$ is a base pair in $R'_1$,
then either $(R'_2[i], R'_2[j])$ is a base pair in $R'_2$
or $R'_2[i] = R'_2[j] = \ '-'$. 
If $(R'_2[i], R'_2[j])$ is a base pair in $R'_2$,
then either $(R'_1[i], R'_1[j])$ is a base pair in $R'_1$
or $R'_1[i] = R'_1[j] = \ '-'$. 
\item 
\label{cond:4}
If $(R'_1[i], R'_1[j])$ and $(R'_1[k], R'_1[l])$ are 
base pairs in $R'_1$ and
$(R'_2[i], R'_2[j])$ and $(R'_2[k], R'_2[l])$ are base pairs
in $R'_2$, then $(R'_1[i], R'_1[j])$ and $(R'_1[k], R'_1[l])$ 
are non-crossing in $R'_1$ and 
$(R'_2[i], R'_2[j])$ and $(R'_2[k], R'_2[l])$ 
are non-crossing in $R'_2$.
\end{enumerate}

\par
From the first three conditions, 
alignments preserve 
the order of unpaired bases and the topological relationship 
between base pairs.
From the last condition, even though the input RNA structures 
may have crossing base pairs, the aligned base pairs are 
non-crossing.

\par
A gap in an alignment $(R'_1, R'_2)$ is a consecutive
subsequences of $'-'$ in either $R'_1$ or $R'_2$ with
maximal length. 
Formally, $[i \cdots j]$ is a gap in $(R'_1, R'_2)$ if
either $R'_1[k] = \ '-'$ for $i \le k \le j$, 
$R'_1[i-1] \ne\ '-'$, and $R'_1[j+1] \ne \ '-'$; 
or $R'_2[k] = \ '-'$ for $i \le k \le j$, 
$R'_2[i-1] \ne \ '-'$, and $R'_2[j+1] \ne \ '-'$.
For each gap in an alignment, in addition to the
insertion/deletion penalty, 
we will assign a constant, $G$, where $G < 0$, 
as the gap penalty. 
This means that longer gaps are preferred since for a longer
gap the additional penalty distributed to each base is 
relatively small.

\par
Given an alignment $(R'_1, R'_2)$, 
we define single-base match $SM$, 
single-base deletion $SD$, 
single-base insertion $SI$,
base-pair match $PM$, 
base-pair deletion $PD$, 
and base-pair insertion $PI$, 
as follows. 
\\ 
\\
$
\begin{array}{l@{\,}l@{\;}l}
{} & SM & = \{i\ |\ R'_1[i] \text{ and } R'_2[i]
\text{ are unpaired bases in }
\\
{} & {} & \qquad \qquad R_1 \text{ and }R_2\}.
\\
{} & SD & = \{i\ |\ R'_1[i] \text{ is an unpaired base in }
R_1 \text{ and}
\\
{} & {} & \qquad \qquad R'_2[i] = \ '-'\}.
\\
{}&SI &=\{i\ |\ R'_2[i]\text{ is an unpaired base in }
R_2 \text{ and}
\\
{} & {} & \qquad \qquad R'_1[i] = \ '-'\}. 
\\
{} & PM & = \{(i,j)\ |\ (R'_1[i], R'_1[j]) \text{ and }
(R'_2[i], R'_2[j])
\\
{} & {} & \qquad \qquad \qquad
\text{are base pairs in }R_1\text{ and }R_2\}.
\\
{} & PD & = \{(i,j)\ |
\ (R'_1[i], R'_1[j])\text{ is a base pair in }
R_1 \text{ and }
\\
{} & {} & \qquad \qquad \qquad R'_2[i] = R'_2[j] = \ '-'\}.
\\
{} & PI & = \{(i, j)\ |
\ (R'_2[i], R'_2[j]) \text{ is a base pair in }
R_2 \text{ and }
\\
{} & {} & \qquad \qquad \qquad R'_1[i] = R'_1[j] = \ '-'\}.
\end{array}
$
\\

\par
The similarity score, $sim((R'_1, R'_2))$, 
of an alignment $(R'_1, R'_2)$ is defined as follows, 
where $N_g$ is the number of gaps in $(R'_1, R'_2)$.
\\
\\
$
\begin{array}{l@{\,}l}
{}&sim((R'_1,R'_2))=
\sum_{i\in SM}\Gamma(R'_1[i]\rightarrow R'_2[i])
\\
{}&\quad + \sum_{i\in SD}\Gamma(R'_1[i]\rightarrow\lambda)
+ \sum_{i\in SI}\Gamma(\lambda\rightarrow R'_2[i])
\\
{}&\quad + \sum_{(i,j)\in PM}\Gamma((R'_1[i],R'_1[j])
\rightarrow (R'_2[i],R'_2[j]))
\\
{}&\quad + \sum_{(i,j)\in PD}\Gamma((R'_1[i],R'_1[j])
\rightarrow\lambda)
\\
{}&\quad + \sum_{(i,j)\in PI}\Gamma(\lambda\rightarrow 
(R'_2[i],R'_2[j])) + G\times N_g.
\end{array}
$
\\
\par
We are now in a position to consider three types of
alignment between two RNA structures $R_1$ and $R_2$. 
The common goal is to find the alignment
with the maximum similarity.
\begin{enumerate}
\item 
Global alignment:
Given two RNA structures $R_1$ and $R_2$, 
find the alignment $(R'_1,R'_2)$ with the maximum similarity
alignment score, 
$
A_G(R_1,R_2)
=\max_{(R'_1,R'_2)}\{sim((R'_1,R'_2))\}.
$
\item 
Local alignment: 
Given two RNA structures $R_1[1\cdots m]$ and $R_2[1\cdots n]$, 
find the alignment 
$
(R'_1[i\cdots j],R'_2[k\cdots l])
$ 
with the maximum similarity alignment score,
$
A_L(R_1,R_2)
=\max_{1\le i\le j\le m\atop 1\le k\le l\le n}
\{sim((R'_1[i\cdots j],R'_2[k\cdots l]))\}.
$
\item 
Pattern matching:
Given two RNA structures 
$
R_1[1\cdots m]$ and $R_2[1\cdots n]
$, 
find the alignment 
$
(R'_1,R'_2[k\cdots l])
$ 
with the maximum similarity alignment score,
$
A_M(R_1,R_2)
=\max_{1\le k\le l\le n}
\{sim((R'_1[1\cdots m],R'_2[k\cdots l]))\}
$.
\end{enumerate}
 
\section{Exact Pattern Matching}
\label{sec:exact}
In this section, we consider the exact RNA structural pattern 
matching problem.
\par
{\bf Exact RNA structural pattern matching}: 
Given a pattern RNA structure $R_1$ and a text RNA structure
$R_2$ of lengths $m$ and $n$ respectively,
determine all the positions where $R_1$ occurs in $R_2$, 
that is,
all positions $i$ in $R_2$ where
\begin{enumerate}
\item 
Primary sequences are identical: 
$R_2[i \cdots i+m-1]=R_1[1 \cdots m]$, 
and
\item 
Structures are identical: 
for any $1 \leq j \leq m$, 
$R_1[j]$ is an
unpaired base if and only if $R_2[i+j-1]$ 
is an unpiared base and for 
any $1 \leq j < k \leq m$, 
$(R_1[j],R_1[k])$ is a base pair if and only if
$(R_2[i+j-1], R_2[i+k-1])$ is a base pair.
\end{enumerate}

\begin{lemma}
\label{lem:exact}
The exact RNA structural pattern matching can be solved 
in $O(n+m)$ time.
\end{lemma}
\emph{Proof}.
Given $R_1$ and $R_2$, 
we first generate $R'_1$ and $R'_2$.
If $R_1[i]$ is an unpaired base, 
then $R'_1[i]=0$ and if $(R_1[i],R_1[j])$ is a
base pair then $R'_1[i]=j-i$ and $R'_1[j]=i-j$.
$R'_2$ is generated in the same manner.
\par
Now we can first perform a standard string matching 
(KMP \cite{KMP77}) to find
all positions where $R_1$ occurs in $R_2$ and then 
perform another
standard string matching to find all positions where 
$R'_1$ occurs in $R'_2$.
These positions where $R_1$ occurs in $R_2$ and $R'_1$ 
occurs in $R'_2$ are
the solution of this problem. 
\indent $\Box$

\section{Approximate Pattern Matching} 
\label{sec:approx}
In this section, 
we consider the approximate RNA structural 
pattern matching problem.
\par
{\bf Approximate RNA structural pattern matching}: 
Given two RNA structures $R_1[1\cdots m]$ and $R_2[1\cdots n]$,
find the best fit of $R_1$ in $R_2$. 
That is,
find the alignment $(R'_1[1\cdots m],R'_2[k\cdots l])$ 
which yields 
$
A_M(R_1,R_2)
=\max_{1\le k\le l\le n}
\{sim((R'_1[1\cdots m],R'_2[k\cdots l]))\}.
$
\par
This problem can be 
viewed as a variation of finding the optimal alignment 
between two RNA structures,
in this case, 
$R_1$ and a part of $R_2$. 
That is, which part
in $R_2$ gives the best alignment to $R_1$? 
This problem 
resembles both global and local alignments in the sense that
we have to match the entire $R_1$ (global) to a part of
$R_2$ (local).
What is presented here, therefore, is an adaptation of an 
algorithm by Wang and Zhang \cite{WangZhang} for
global alignment between two RNA structures taking affine gap
penalty into consideration. 
We modify the algorithm in combination with a known algorithm
for fitting one sequence into another \cite{Waterman} to 
yield an algorithm for approximate pattern 
matching with affine gap penalty between two RNA structures.
\par
The algorithm is based on dynamic programming and uses a 
bottom-up procedure consisting of two phases. 
In the first phase, we compute global alignments between 
substructures in $R_1$ and substructures in $R_2$. 
In the second phase, we take the two 
whole structures $R_1$ and $R_2$ and compute pattern matching.

\subsection{Properties} 
\label{sec:approx-prop}
In the original model \cite{WangZhang},
a distance measure is used. For the approximate
algorithm in this paper, we convert the distance measure
to the similarity measure.
\par
Consider two RNA structures $R_1$ and $R_2$.
We use $g$, where $g > 0$, 
to denote the gap penalty.
We use $\gamma(i,j)$, where $1 \le i \le |R_1|$ and
$1 \le j \le |R_2|$, to denote the similarity score 
for aligning two structural elements $e_i \in S(R_1)$
and $e_j \in S(R_2)$
(c.f., section \ref{sec:rep}).
Note that $\gamma(i,j)$ is positive if the two structural
elements match, and negative otherwise.
We use $\delta(i,0)$ and $\delta(0,j)$, 
where $1 \le i \le |R_1|$ and $1 \le j \le |R_2|$, to denote 
the penalty of aligning 
$R_1[i]$ and $R_2[j]$, respectively, to $'-'$.
Note that $\delta(i,0) \text{ and } \delta(0,j) > 0$.
We define $\gamma$ and $\delta$ in terms of $\Gamma$ as follows.
\\ 
\\
$
\begin{array}{l@{\,}l}
{}&\delta(i,0)=-\Gamma(R_1[i]\rightarrow\lambda)\\
{}&\quad \text{if }i=p(R_1[i]), \\
{}&\delta(0,j)=-\Gamma(\lambda\rightarrow R_2[j])\\
{}&\quad \text{if }j=p(R_2[j]), \\
{}&\gamma(i,j)=\Gamma(R_1[i]\rightarrow R_2[j])\\
{}&\quad \text{if }i=p(R_1[i])\text{ and }j=p(R_2[j]), \\
{}&\delta(i,0)=\delta(i',0)
=-\Gamma((R_1[i],R_1[i'])\rightarrow\lambda)/2 \\
{}&\quad \text{if }i=p(R_1[i'])<i'\le |R_1|,\\
{}&\delta(0,j)=\delta(0,j')
=-\Gamma(\lambda\rightarrow (R_2[j],R_2[j']))/2 \\
{}&\quad \text{if }j=p(R_2[j'])<j'\le |R_2|,\\
{}&\gamma(i,j)=\Gamma((R_1[i'],R_1[i])\rightarrow 
(R_2[j'],R_2[j])) \\
{}&\quad \text{if }
0<i' =p(R_1[i])<i \text{ and }
0<j' =p(R_2[j])<j.
\end{array}
$
\\
\par
Consider the optimal-alignment score between 
$R_1[i_1\cdots i_2]$ and $R_2[j_1\cdots j_2]$. 
We use $A(i_1\cdots i_2,j_1\cdots j_2)$ to denote the 
optimal-alignment 
score between $R_1[i_1\cdots i_2]$ and $R_2[j_1\cdots j_2]$,
$D(i_1\cdots i_2,j_1\cdots j_2)$ the optimal-alignment 
score between $R_1[i_1\cdots i_2]$ and $R_2[j_1\cdots j_2]$ 
such that the alignment ends with $R_1[i_2]$ aligned to $'-'$, 
and
$I(i_1\cdots i_2,j_1\cdots j_2)$ the optimal-alignment 
score between $R_1[i_1\cdots i_2]$ 
and $R_2[j_1\cdots j_2]$ such that the alignment ends with
$R_2[j_2]$ aligned to $'-'$.
\par
In the lemmas that follow, we attach an index to 
$A$, $D$ and $I$ to indicate the phase of
execution.
\par
As most of the lemmas are proved in \cite{WangZhang},
we give the proofs pertaining to the modified lemmas
with respect to those therein. 

\subsubsection{Phase 1} 
\label{sec:approx-prop-phase1}

\begin{lemma} 
\label{lem:ap:A}
\textnormal{\cite{WangZhang}} 
\\*
\indent
$
\begin{array}{l@{\;=\;}l@{\quad}l@{\;=\;}l@{\quad}l@{\;=\;}l}
A_1\left(\emptyset,\emptyset\right)&0,&
D_1\left(\emptyset,\emptyset\right)&-g,&
I_1\left(\emptyset,\emptyset\right)&-g.
\end{array}
$
\end{lemma}
\emph{Proof}.
For $A_1(\emptyset, \emptyset)$, consider $A_1(i_1, j_1)$
when the optimal alignment results from aligning
$R_1[i_1]$ to $R_2[j_1]$, in which case 
$A_1(i_1, j_1) = \gamma(i_1,j_1)$. In comparison with 
the first case
in lemma \ref{lem:ap:E}, we may set 
$A_1(\emptyset, \emptyset) = 0$.
\par
For $D_1(\emptyset, \emptyset)$, consider
$D_1(i_1, \emptyset)$ by which $R_1[i_1]$ is aligned to
$'-'$ and a gap is opened, hence 
$D_1(\emptyset, \emptyset) = -g$. 
Similarly, we may obtain that
$I_1(\emptyset, \emptyset) = -g$.
Moreover, if we substitute the values of 
$A_1(\emptyset, \emptyset)$ and $D_1(\emptyset, \emptyset)$
into lemma \ref{lem:ap:C} considering $D_1(i_1, \emptyset)$, 
we have the same value for both possibilities. This is true
since there is only one way to align $R_1[i_1]$ to $'-'$, hence
only one possible value to select. 
\indent $\Box$

\begin{lemma} 
\label{lem:ap:B}
\textnormal{\cite{WangZhang}} 
$For\ i_1\le i\le i_2\ and\ j_1\le j\le j_2$,
\\*
$
\begin{array}{l@{\;=\;}l}
D_1\left(i_1\cdots i,\emptyset\right)&
D_1\left(i_1\cdots i-1,\emptyset\right)
-\delta\left(i,0\right),
\\ 
A_1\left(i_1\cdots i,\emptyset\right)&
D_1\left(i_1\cdots i,\emptyset\right),
\\ 
I_1\left(i_1\cdots i,\emptyset\right)&
D_1\left(i_1\cdots i,\emptyset\right)
-g,
\\ 
I_1\left(\emptyset,j_1\cdots j\right)&
I_1\left(\emptyset,j_1\cdots j-1\right)
-\delta\left(0,j\right),
\\ 
A_1\left(\emptyset,j_1\cdots j\right)&
I_1\left(\emptyset,j_1\cdots j\right),
\\
D_1\left(\emptyset,j_1\cdots j\right)&
I_1\left(\emptyset,j_1\cdots j\right)
-g.
\end{array}
$
\end{lemma}
\emph{Proof}.
For $D_1(i_1 \cdots i, \emptyset)$, by definition $R_1[i]$
is aligned to $'-'$, hence the $-\delta(i,0)$ term;
and $R_1[i_1 \cdots i-1]$ is aligned to $\emptyset$.
There is only one
possibility, namely, each element in $R_1[i_1 \cdots i-1]$ is
aligned to $'-'$, by which we know that $R_1[i-1]$ is also 
aligned to $'-'$, hence by definition the
$D_1(i_1 \cdots i-1, \emptyset)$ term. 
\par
For $A_1(i_1 \cdots i, \emptyset)$, this is the optimal alignment
between $R_1[i_1 \cdots i]$ and $\emptyset$. 
There is only one
possibility, namely, each element in $R_1[i_1 \cdots i]$ is
aligned to $'-'$, by which we know that $R_1[i]$ is also 
aligned to $'-'$, hence by definition 
$A_1(i_1 \cdots i, \emptyset) = D_1(i_1 \cdots i, \emptyset)$.
\par
For $I_1(i_1 \cdots i, \emptyset)$, consider
$I_1(i_1 \cdots i, j_1)$, the optimal
alignment between $R_1[i_1 \cdots i]$ and 
$R_2[j_1]$ that ends with $R_2[j_1]$ aligned to $'-'$.
This yields only one possible alignment, namely,
$R_2[j_1]$ is aligned to $'-'$ and each element in
$R_1[i_1 \cdots i-1]$ is aligned to $'-'$.
Therefore, both possibilities in lemma \ref{lem:ap:D}
should yield the same value for $I_1(i_1 \cdots i, j_1)$. 
That is,
$I_1(i_1 \cdots i, \emptyset) - \delta(0,j) = 
A_1(i_1 \cdots i, \emptyset) - \delta(0,j) -g$.
We have just shown that 
$A_1(i_1 \cdots i, \emptyset) = 
D_1(i_1 \cdots i, \emptyset)$, hence
$I_1(i_1 \cdots i, \emptyset) = D_1(i_1 \cdots i, \emptyset) - g$.
Meanwhile, we may think that aligning 
$R_2[j_1]$ to $'-'$
opens a gap, hence the $-g$ term. Then there is
only one way to optimally align $R_1[i_1 \cdots i]$
to $\emptyset$, namely, aligning each element in
$R_1[i_1 \cdots i]$ to $'-'$, which forces $R_1[i]$
to be aligned to $'-'$, hence the 
$D_1(i_1 \cdots i, \emptyset)$ term. 
If $I_1(i_1 \cdots i, j_1)$ comes from 
$I_1(i_1 \cdots i, \emptyset)$, then
$I_1(i_1 \cdots i, \emptyset)$ must account for both terms.
Applying same line of arguments for 
$D_1(i_1 \cdots i, \emptyset)$,
$A_1(i_1 \cdots i, \emptyset)$ and
$I_1(i_1 \cdots i, \emptyset)$
yields the corresponding formulae for 
$I_1(\emptyset, j_1 \cdots j)$,
$A_1(\emptyset, j_1 \cdots j)$ and
$D_1(\emptyset, j_1 \cdots j)$, respectively.
\indent $\Box$

\begin{lemma}
\label{lem:ap:C}
\textnormal{\cite{WangZhang}} 
$
For\ i_1\le i\le i_2\ and\ j_1\le j\le j_2,
\\*
D_1\left(i_1\cdots i,j_1\cdots j\right)
\\
= max 
\begin{cases}
D_1\left(i_1\cdots i-1,j_1\cdots j\right)-
\delta\left(i,0\right), 
\\
A_1\left(i_1\cdots i-1,j_1\cdots j\right)-
\delta\left(i,0\right)
-g.
\end{cases}
$
\end{lemma}

\begin{lemma} 
\label{lem:ap:D}
\textnormal{\cite{WangZhang}} 
$
For\ i_1\le i\le i_2\ and\ j_1\le j\le j_2,
\\*
I_1\left(i_1\cdots i,j_1\cdots j\right)
\\
= max
\begin{cases}
I_1\left(i_1\cdots i,j_1\cdots j-1\right)
-\delta\left(0,j\right), 
\\
A_1\left(i_1\cdots i,j_1\cdots j-1\right)
-\delta\left(0,j\right)
-g.
\end{cases}
$
\end{lemma}

\begin{lemma} 
\label{lem:ap:E}
\textnormal{\cite{WangZhang}} 
$For\ i_1\le i\le i_2\ and\ j_1\le j\le j_2,
\\*
if\ i=p(R_1[i])\ and\ j=p(R_2[j]),\ then
\\ 
A_1\left(i_1\cdots i,j_1\cdots j\right)
\\
= max 
\begin{cases}
D_1\left(i_1\cdots i,j_1\cdots j\right),
\\
I_1\left(i_1\cdots i,j_1\cdots j\right),
\\
A_1\left(i_1\cdots i-1,j_1\cdots j-1\right)
+\gamma\left(i,j\right);
\end{cases}
\\
if\ i_1\le p(R_1[i])<i\ and\ j_1\le p(R_2[j])<j,\ then
\\
A_1\left(i_1\cdots i,j_1\cdots j\right)= max
\\
\begin{cases}
D_1\left(i_1\cdots i,j_1\cdots j\right),
\\
I_1\left(i_1\cdots i,j_1\cdots j\right),
\\
A_1\left(i_1\cdots p\left(R_1[i]\right)-1,
j_1\cdots p\left(R_2[j]\right)-1\right)
\\
{}+A_1\left(p\left(R_1[i]\right)+1\cdots i-1,
p\left(R_2[j]\right)+1\cdots j-1\right)
\\
{}+\gamma\left(i,j\right);
\end{cases}
\\
otherwise,
\\
A_1\left(i_1\cdots i,j_1\cdots j\right)= max 
\begin{cases}
D_1\left(i_1\cdots i,j_1\cdots j\right),
\\
I_1\left(i_1\cdots i,j_1\cdots j\right).
\end{cases}$
\end{lemma}

\subsubsection{Phase 2} 
\label{sec:approx-prop-phase2}
The lemmas in phase 1 are for global alignment. 
For our interested problem, 
namely finding the best fit of $R_1$ in $R_2$, 
we need to modify some of those lemmas.
\par
The major changes occur in the boundary conditions, 
i.e., 
lemmas \ref{lem:ap:a} and \ref{lem:ap:b}.
The idea for the boundary conditions is the same 
as that in the sequence case \cite{Waterman}. 
That is, initial alignments to $'-'$ in $R_2$ are
not accounted for as penalties until the first match
is met; each alignment to $'-'$ thereafter in $R_2$
accounts for a penalty. For $R_1$, however, each
alignment to $'-'$ accounts for a penalty.
\par
Additionally in case 2 of 
lemma \ref{lem:ap:e}, an $A_1$ term is used
because it corresponds to an alignment between two 
substructures, each delimited by a base pair, that has
been computed in phase 1.
\par
Other than the above-mentioned changes, the lemmas in 
phase 2 are the same as those in phase 1. 

\begin{lemma} 
\label{lem:ap:a}
\ \\*
\indent
$
\begin{array}{l@{\;=\;}l@{\quad}l@{\;=\;}l}
A_2\left(\emptyset,\emptyset\right)&0,& 
D_2\left(\emptyset,\emptyset\right)&-g.
\end{array}
$
\end{lemma}
\emph{Proof}.
Since this is a subset of lemma \ref{lem:ap:A}, same
arguments hold.
\indent $\Box$

\begin{lemma} \label{lem:ap:b}
$For\ 1\le i\le |R_1|\ and\ 1\le j\le |R_2|$,
\\*
$
\begin{array}{l@{\;=\;}l}
D_2\left(1\cdots i,\emptyset\right)&
D_2\left(1\cdots i-1,\emptyset\right)
-\delta\left(i,0\right),
\\ 
A_2\left(1\cdots i,\emptyset\right)&
D_2\left(1\cdots i,\emptyset\right),
\\ 
I_2\left(1\cdots i,\emptyset\right)&
D_2\left(1\cdots i,\emptyset\right)
-g,
\\ 
A_2\left(\emptyset,1\cdots j\right)&0,
\\
D_2\left(\emptyset,1\cdots j\right)&-g.
\end{array}
$
\end{lemma}
\emph{Proof}.
We prove only the modified part with respect to 
lemma \ref{lem:ap:B}.
\par
For $A_2(\emptyset, 1 \cdots j)$, the only possibility
is to align each element in $R_2(1 \cdots j)$ to
$'-'$. In finding the best fit of a pattern in $R_2$,
each initial alignment to $'-'$ in $R_2$ before
the first match is not accounted for as penalty, hence
$A_2(\emptyset, 1 \cdots j) = 0$.
\par
For $D_2(\emptyset, 1 \cdots j)$, consider 
$D_2(1, 1 \cdots j)$ when $R_1[1]$ is aligned to $'-'$,
which means that a gap is opened, hence 
$D_2(\emptyset, 1 \cdots j)$ accounts for the $-g$ term.
Consequently, the entire $R_2[1 \cdots j]$ can only be
aligned to $'-'$. As just mentioned, this does not
invoke a penalty, hence no other terms are accounted for 
by $D_2(\emptyset, 1 \cdots j)$. 
\indent $\Box$

\begin{lemma} 
\label{lem:ap:c}
\textnormal{\cite{WangZhang}} 
$For\ 1\le i\le |R_1|\ and\ 1\le j\le |R_2|,
\\*
D_2\left(1\cdots i,1\cdots j\right)
\\
= max 
\begin{cases}
D_2\left(1\cdots i-1,1\cdots j\right)
-\delta\left(i,0\right),
\\
A_2\left(1\cdots i-1,1\cdots j\right)
-\delta\left(i,0\right)-g.
\end{cases}
$
\end{lemma}

\begin{lemma} 
\label{lem:ap:d}
\textnormal{\cite{WangZhang}} 
$For\ 1\le i\le |R_1|\ and\ 1\le j\le |R_2|,
\\*
I_2\left(1\cdots i,1\cdots j\right)
\\
= max
\begin{cases}
I_2\left(1\cdots i,1\cdots j-1\right)
-\delta\left(0,j\right),
\\
A_2\left(1\cdots i,1\cdots j-1\right)
-\delta\left(0,j\right)-g.
\end{cases}
$
\end{lemma}

\begin{lemma} 
\label{lem:ap:e}
\textnormal{\cite{WangZhang}} 
$For\ 1\le i\le |R_1|\ and\ 1\le j\le |R_2|,
\\*
if\ i=p(R_1[i])\ and\ j=p(R_2[j]),\ then
\\ 
A_2\left(1\cdots i,1\cdots j\right)
\\
= max
\begin{cases}
D_2\left(1\cdots i,1\cdots j\right),
\\
I_2\left(1\cdots i,1\cdots j\right),
\\
A_2\left(1\cdots i-1,1\cdots j-1\right)
+\gamma\left(i,j\right);
\end{cases}
\\
if\ 1\le p(R_1[i])<i\ and\ 1\le p(R_2[j])<j,\ then
\\
A_2\left(1\cdots i,1\cdots j\right)= max
\\
\begin{cases}
D_2\left(1\cdots i,1\cdots j\right),
\\
I_2\left(1\cdots i,1\cdots j\right),
\\
A_2\left(1\cdots p\left(R_1[i]\right)-1,
1\cdots p\left(R_2[j]\right)-1\right)
\\
{}+A_1\left(p\left(R_1[i]\right)+1\cdots i-1,
p\left(R_2[j]\right)+1\cdots j-1\right)
\\
{}+\gamma\left(i,j\right);
\end{cases}
\\
otherwise,
\\
A_2\left(1\cdots i,1\cdots j\right)= max
\begin{cases}
D_2\left(1\cdots i,1\cdots j\right),
\\
I_2\left(1\cdots i,1\cdots j\right).
\end{cases}$
\end{lemma}

\subsection{Algorithm} 
\label{sec:approx-alg}
We use a bottom-up dynamic-programming algorithm executing
two phases described as follows.

\subsubsection{Phase 1} 
\label{sec:approx-alg-phase1}
In this phase, we compute alignments between substructures
as follows. 
We compute the alignments between every pair of substructures 
$R_1\left[i_1+1\cdots i_2-1\right]$ and 
$R_2\left[j_1+1\cdots j_2-1\right]$,
where $p(R_1[i_1])=i_2$ and $p(R_2[j_1])=j_2$ 
(i.e., $i_1$ and $i_2$ form a base pair in $R_1$; ditto 
for $j_1$ and $j_2$ in $R_2$).
For nested pairs, we start with the innermost pair then
proceed outwards.
\par
Specifically, given $R_1$ and $R_2$, we first sort 
$P(R_1)$ and $P(R_2)$ by 3' end
into two sorted lists $L_1$ and $L_2$, respectively. 
For each pair of base pairs $(L_1[i],L_2[j])$, where
$L_1[i]=(i_1, i_2)$ and $L_2[j]=(j_1, j_2)$,
we use lemmas \ref{lem:ap:A} to \ref{lem:ap:E} to compute
$A_1(i_1+1\cdots i_2-1,j_1+1\cdots j_2-1)$.

\subsubsection{Phase 2} 
\label{sec:approx-alg-phase2}
In this phase, we take the two whole structures,
$R_1$ and $R_2$, and compute
$A_2(R_1,R_2)=A_2(1\cdots |R_1|,1\cdots |R_2|)$
using lemmas \ref{lem:ap:a} to \ref{lem:ap:e}.

\subsubsection{Finding the Best Fit} 
\label{sec:approx-alg-findfit}
To find the best fit involving
$R_1[1\cdots m]$ and $R_2[k\cdots l]$,
where $m=|R_1|$, we can do a 
traceback starting at $A_{2_{m,l}}(R_1,R_2)$, where
the subscript $(m,l)$ refers to the entry position
in the associated matrix. Note that traceback
may involve multiple matrices.

\subsubsection{Complexities} 
\label{sec:approx-alg-compl}
The running time to compute $A_i\left(i_1\cdots i_2,j_1\cdots j_2\right)$,
$D_i\left(i_1\cdots i_2,j_1\cdots j_2\right)$ and
$I_i\left(i_1\cdots i_2,j_1\cdots j_2\right)$,
where $i = 1\text{ or }2$,
is bounded by $O(|R_1||R_2|)$.  
Let $P_1$ and $P_2$ be the numbers of base pairs
in $R_1$ and $R_2$, respectively.
The worst-case time complexity of the algorithm is 
$O(P_1P_2|R_1||R_2|)$, which can be improved
to $O(S_1S_2|R_1||R_2|)$ where $S_1$ and $S_2$ are the numbers
of stems in $R_1$ and $R_2$, respectively.
The space complexity of the algorithm is $O(|R_1||R_2|)$.

\section{Local Alignment} 
\label{sec:local}
In this section, we consider the local structural alignment
problem. The ideas follow directly from the sequence case
\cite{Waterman}. With respect to the structural case, it is
a simple extension from the approximate structural pattern
matching problem in section \ref{sec:approx}. 
We also use a two-phase procedure much the
same as presented in section \ref{sec:approx}. 
The lemmas for phase 1 are the same as those in phase 1
of section \ref{sec:approx}. We list only the lemmas for
phase 2. The modifications are trivial. Same proof reasonings
in section \ref{sec:approx} can easily apply here. Due to page 
limit, we omit the proofs. We remark two major changes. 
First, the boundary conditions. Essentially, if we extend
half of the boundary conditions of the approximate pattern
matching problem to the other half, we get the boundary 
conditions for the local alignment problem. This is easy to see
since the boundary conditions for the text structure behave
just like local alignment. Secondly, in computing the
optimal-alignment score, an additional possibility is included,
namely 0.

\begin{lemma} 
\label{lem:lo:a}
\ \\*
\indent \indent \indent
$
\begin{array}{l@{\;=\;}l}
A_2\left(\emptyset,\emptyset\right)&0. 
\end{array}
$
\end{lemma}

\begin{lemma} 
\label{lem:lo:b}
$For\ 1\le i\le |R_1|\ and\ 1\le j\le |R_2|$,
\\*
\indent
$
\begin{array}{l@{\;=\;}l@{\quad}l@{\;=\;}l}
A_2\left(1\cdots i,\emptyset\right)&0,&
I_2\left(1\cdots i,\emptyset\right)&-g,
\\ 
A_2\left(\emptyset,1\cdots j\right)&0,&
D_2\left(\emptyset,1\cdots j\right)&-g.
\end{array}
$
\end{lemma}

\begin{lemma} 
\label{lem:lo:c}
$For\ 1\le i\le |R_1|\ and\ 1\le j\le |R_2|,
\\*
D_2\left(1\cdots i,1\cdots j\right)
\\
= max 
\begin{cases}
0,
\\
D_2\left(1\cdots i-1,1\cdots j\right)
-\delta\left(i,0\right),
\\
A_2\left(1\cdots i-1,1\cdots j\right)
-\delta\left(i,0\right)-g.
\end{cases}
$
\end{lemma}

\begin{lemma} 
\label{lem:lo:d}
$For\ 1\le i\le |R_1|\ and\ 1\le j\le |R_2|,
\\*
I_2\left(1\cdots i,1\cdots j\right)
\\
= max
\begin{cases}
0,
\\
I_2\left(1\cdots i,1\cdots j-1\right)
-\delta\left(0,j\right),
\\
A_2\left(1\cdots i,1\cdots j-1\right)
-\delta\left(0,j\right)-g.
\end{cases}
$
\end{lemma}

\begin{lemma} 
\label{lem:lo:e}
$For\ 1\le i\le |R_1|\ and\ 1\le j\le |R_2|,
\\*
if\ i=p(R_1[i])\ and\ j=p(R_2[j]),\ then
\\ 
A_2\left(1\cdots i,1\cdots j\right)
\\
= max
\begin{cases}
0,
\\
D_2\left(1\cdots i,1\cdots j\right),
\\
I_2\left(1\cdots i,1\cdots j\right),
\\
A_2\left(1\cdots i-1,1\cdots j-1\right)
+\gamma\left(i,j\right);
\end{cases}
\\
if\ 1\le p(R_1[i])<i\ and\ 1\le p(R_2[j])<j,\ then
\\
A_2\left(1\cdots i,1\cdots j\right)= max
\\
\begin{cases}
0,
\\
D_2\left(1\cdots i,1\cdots j\right),
\\
I_2\left(1\cdots i,1\cdots j\right),
\\
A_2\left(1\cdots p\left(R_1[i]\right)-1,
1\cdots p\left(R_2[j]\right)-1\right)
\\
{}+A_1\left(p\left(R_1[i]\right)+1\cdots i-1,
p\left(R_2[j]\right)+1\cdots j-1\right)
\\
{}+\gamma\left(i,j\right);
\end{cases}
\\
otherwise,
\\
A_2\left(1\cdots i,1\cdots j\right)= max
\begin{cases}
0,
\\
D_2\left(1\cdots i,1\cdots j\right),
\\
I_2\left(1\cdots i,1\cdots j\right).
\end{cases}$
\end{lemma}

\section{Conclusions} 
\label{sec:conclusions}
We have presented three algorithms, of which, two pertain to
the exact and approximate RNA structural pattern matching 
problems; and one, the local RNA structural alignment
problem. 
\par
In the exact pattern-matching algorithm, 
we use a labelling device to
preserve the relative structural information within
a given RNA structure, thereby
facilitating direct use of an existing fast string-matching
algorithm on the exact RNA structural pattern matching problem.
\par
In the approximate pattern-matching algorithm, 
we adapt an earlier work
in RNA structural alignment to the approximate RNA structural 
pattern matching problem.
\par
We further extend the approximate pattern-matching algorithm
to adapt to the local RNA structural alignment problem.

\clearpage
\end{multicols}
\end{document}